\documentclass{aa}  

\usepackage{graphicx}
\usepackage{mathrsfs}
\usepackage{txfonts}
\begin{document} 

\title{GU~Mon, a high-mass eclipsing overcontact binary in the young open cluster Dolidze~25}

   \author{J. Lorenzo \inst{1}
\and I. Negueruela \inst{1}
\and F. Vilardell \inst{2}
\and S. Sim\'on-D\'{\i}az \inst{3,4}
\and P. Pastor \inst{5} 
\and M. M\'endez Majuelos \inst{6}}

   \institute{Departamento de F\'{\i}sica, Ingenier\'{\i}a de Sistemas y 
Teor\'{\i}a de la Se\~nal, Escuela Polit\'ecnica Superior, Universidad de Alicante, Carretera de San Vicente del Raspeig s/n, E03690, San Vicente del Raspeig, Alicante, Spain
   \and Institut d’Estudis Espacials de Catalunya, Edifici Nexus, c/ Capit\'a, 2-4, desp. 201, E08034 Barcelona, Spain
   \and
Instituto de Astrof\'{\i}sica de Canarias,  V\'ia L\'actea s/n, E38200, La Laguna, Tenerife, Spain 
\and
Departamento de Astrof\'{\i}sica, Universidad de La Laguna, Facultad de F\'isica y Matem\'aticas, Universidad de La Laguna, Avda. Astrof\'isico Francisco S\'anchez, s/n, E38205, La Laguna, Tenerife, Spain
\and
Departamento de Lenguajes y Sistemas Inform\'{a}ticos, Universidad de Alicante, Apdo. 99, E03080 Alicante, Spain 
\and 
Departamento de Ciencias, IES Arroyo Hondo, c/ Maestro Manuel Casal 2, E11520, Rota, C\'adiz, Spain }
   \date{Received ; accepted }

 
  \abstract
   {The eclipsing binary GU~Mon is located in the star-forming cluster Dolidze~25, which has the lowest metallicity 
   measured in a Milky Way young cluster.}
   {GU~Mon has been identified as a short-period eclipsing binary with two early B-type components. We set out to 
   derive its orbital and stellar parameters.} 
{We present a comprehensive analysis, including $B$ and $V$ light curves and 11 high-resolution spectra, to verify 
the orbital period and determine parameters. We use the stellar atmosphere code {\sc fastwind} to obtain stellar 
parameters and create templates for cross-correlation. We obtain a model
   to fit the light and radial-velocity curves using the Wilson-Devinney code iteratively and simultaneously.}
{The two components of GU~Mon are identical stars of spectral type B1\,V, with the same mass and temperature. 
The lightcurves are typical of an EW-type binary. The spectroscopic and photometric analyses agree on a period 
of $0.896640 \pm 0.000007$~d. We determine a mass of $9.0 \pm 0.6\:M_{\sun}$ for each component and temperatures of 
$28\,000 \pm 2\,000$~K. Both values are consistent with the spectral type. The two stars are overfilling their respective Roche lobes, 
sharing a common envelope, and therefore the orbit is synchronised and circularised. }
   {The GU~Mon system has a fill-out factor above 0.8, containing two dwarf B-type stars on the main sequence. 
   The two stars are in a very advanced stage of interaction, with their extreme physical similarity likely due 
   to the common envelope. The expected evolution of such a system will very probably lead to a merger while still on the main sequence.}


   \maketitle
%

\section{Introduction}

Understanding the effects of binary interaction on the evolution of high-mass stars has become a very active topic of research. The relationship 
between fast rotation and binary interactions is widely discussed \citep[e.g.][]{lang2012}, since it can perhaps explain the observed distribution of rotational velocities among O-type stars \citep{simo2014}.

One of the possible consequences of binary interaction is the formation of high-mass stars via a merger channel. Mergers have been considered a possible mechanism to form massive stars. However, direct collisions of stars in open clusters require extremely dense clusters \citep[$\approx 10^8$ stars pc$^{-3}$;][]{clark2006}. If the merger happens in a binary, the stellar density required to induce mergers is $\approx 10^6$ stars pc$^{-3}$  \citep{bonn2005},
two orders of magnitude smaller than the stellar density with direct collisions. On the other hand, the binary
fraction in young clusters can achieve percentages of 75\% \citep{hu2007, sana2012}. Thus, it is not difficult to envisage a scenario for high-mass stars forming as a result of binary systems whose components end up merging. Mergers are more likely to occur in dense stellar environments, where dynamical perturbations may induce them \citep[e.g.][]{bane2012}. 

Recently, \cite{demi2014} simulated populations of high-mass stars, assuming a constant star formation rate. They found 
that the products of mergers resulting from close binary systems account for 8$^{+9}_{-4}\%$  of a sample of early-type stars. In total, $30^{+10}_{-15}\%$ of high-mass main-sequence stars are the products of binary interactions \citep{demi2014}. Observationally, the binary star MY~Camelopardalis, a member of the very young 
cluster Alicante~1 \citep{negu2008}, has been shown to be composed by two rather massive stars ($M_{*}\ga30\,M_{\sun}$) in a close orbit,
with 
orbital parameters such that it should merge according to standard binary evolution models \citep{lore2014}.

Another possible consequence of binarity is the formation of blue stragglers in young open clusters. The validity of the coalescence of a binary system as a mechanism to form blue stragglers has been discussed in several works \citep[e.g.][]{zinn1976}. This kind of mechanism is generally studied with regard to low-mass stars \citep[e.g.][]{leon1992, becc2013}. However, blue stragglers in young open clusters may also be related to binary interactions \citep{marc2007}. There are two possible scenarios, depending on the initial binary separation. A star may become a blue straggler because it accretes material from a companion, thus rejuvenating \citep{well1999}. The companion then continues its evolution, leading to the possibility of a second binary interaction, unless the system is disrupted in a supernova explosion. The other option is the merger of two high-mass stars, if the initial separation is sufficiently small, as discussed above.

In this paper we analyse the properties of a probable progenitor to a high-mass merger, GU~Mon.
GU~Mon is an eclipsing binary located in Dolidze~25, a very young open cluster in the Galactic Anticentre associated with the 
distant \ion{H}{ii} region Sh-2~284 \citep{shar1959}. Its galactic coordinates are $l=212\fdg0$ and $b=-1\fdg4$. In a recent paper (\citealt{negu2015}, henceforth Paper~I), we derive a distance of 4.5~kpc to 
the cluster, showing that its stars present very low abundances of oxygen and silicon (about 0.4~dex below the solar value, 
making Dolidze~25 the young open cluster with the lowest metallicity known in the Milky Way). GU~Mon is star 13 in Dolidze~25, 
according to the numbering system of \citet{moff1975}, who identify it as an early-type star. Using a very-low-dispersion 
spectrogram, \citet{babu1983} classified GU~Mon as B8\,V with an apparent magnitude of $V=12.3$. Later, \citet{lenn1990} 
took a spectrum of the star and identified it as an early-B star with broad lines. GU~Mon was catalogued as an eclipsing binary
of EW type by \citet{zejd2002}, who derived a period of 0.89668149~d. \cite{krei2004} revised the
ephemeris for a wide collection of eclipsing binaries,  finding for GU~Mon a period of 0.8966485~d and HJD~2452500.5891 as time of minimum.
\citet{hubs2005} calculated new times of minima, using photoelectric photometry, 950 cycles after the previous minimum 
(HJD~2453352.4110 in the heliocentric system). GU~Mon lies within the field of view of the CoRoT space mission. Its photometric 
lightcurve was analysed by \citet{maci2011}, who derived a period of 0.897~d, an inclination of the orbit of $72\fdg137$ and mass 
ratio of $0.9$, under the hypothesis that the two components had spectral type F2\,II. The choice of this spectral type, at odds 
with all previous determinations, is not explained, and likely based on the assumption that the star is not reddened. 

In the following, we present a comprehensive analysis of GU~Mon, deriving its orbital and physical parameters. GU~Mon represents an 
example of a high-mass binary system in a young open cluster whose components are already in a common envelope. Due to the fraction of the Roche lobe already shared, both stars will in all likelihood merge. We suggest that GU~Mon 
represents a crucial stage in the evolution of such systems,
presenting a \textquotedblleft peanut-shaped \textquotedblright surface.
The paper is arranged as follows: We first describe the observations
consisting of a spectroscopic survey and photometric data. Then we carry out a comprehensive spectroscopic analysis,
including spectral classification, morphological description of the spectrum, determination of  the radial velocity and spectrum modelling. After this, we analyse the radial velocity and light curves, and show the results of the orbital parameters obtained from the Wilson-Devinney code. Finally,
we discuss the evolutionary state of GU~Mon and present our conclusions.

\defcitealias{negu2015}{Paper~I}


\section{Observations}

We carried out the spectroscopic observations with the high-resolution FIbre-fed Echelle Spectrograph (FIES) 
attached to the Nordic Optical Telescope (NOT), located at the Observatorio del Roque de los Muchachos (La Palma, Spain) 
between 2011 January 13 and 16. Eleven spectra were 
registered with exposure times of 1800~s. The signal-to noise ratio achieved 
is between 30 and~50. We selected the low-resolution mode (R=25\,000), which covers the spectral range between 3700 and 7300\,\AA\, without gaps in a single, fixed 
setting \citep{telt2014}. 
 The spectra were homogeneously reduced using the
FIEStool\footnote{http://www.not.iac.es/instruments/fies/fiestool/FIEStool.html} software in advanced mode. A complete 
set of bias, flat and arc frames, obtained each night, was used to this aim.  For wavelength calibration, we used arc 
spectra of a ThAr lamp. Heliocentric corrections were calculated using the {\sc rv} program included in the {\it Starlink} suite. 
We list the HJD, date, and phase for every spectrum in Tab.~\ref{log}. Spectra are numbered and sorted by date in ascending order. 

\begin{table}[!ht]
\caption{Log of spectroscopic observations. The spectrum $\sharp$12 was obtained with ISIS spectrograph,
described in the text, but not included in the analysis.\label{log}}
\centering 
\scalebox{0.94}{
\begin{tabular}{c c c c}
\hline \hline
\noalign{\smallskip}
$\sharp$&HJD&UT date&signal-to-noise\\
&-2450000&year-month-day& ratio\\
\noalign{\smallskip}
\hline
\noalign{\smallskip}
     1	&5574.52029&2011-01-13&45\\
     2	&5574.54390&2011-01-13&47\\
     3	&5574.59112&2011-01-13&38\\
     4	&5574.60709&2011-01-13&38\\
     5	&5575.38001&2011-01-13&43\\
     6	&5575.45015&2011-01-13&45\\
     7	&5575.52237&2011-01-14&40\\
     8	&5575.61195&2011-01-14&30\\
     9	&5576.37931&2011-01-14&32\\
    10	&5577.64320&2011-01-16&37\\
    11	&5577.65778&2011-01-16&36\\
    12&5870.62462&2011-11-05&150\\
    13&5874.56900&2011-11-09&150\\
\noalign{\smallskip}
\hline
\end{tabular}}

\end{table}

In addition, lower-resolution spectroscopy was obtained with the ISIS double-beam spectrograph, mounted on the 4.2~m 
William Herschel Telescope (WHT) in La Palma (Spain). The observations were taken in service mode on the nights of November 2011, 
4th and 8th, with exposure time of 900\,s ($2\times450$~s poses). The blue arm of the instrument was equipped with the R1200B grating and the EEV12 CCD. This configuration covers an unvignetted range of $\approx650$\AA\, with a nominal dispersion of 0.23\:\AA/pixel. To observe the whole classification range, as displayed in Fig.~\ref{spec}, two exposures were taken, centred on 4250 and 4900\:\AA. With a $1\farcs5$ slit, the resolution element is about 1.2\:\AA{}, giving a resolving power in the classification region $R\sim4\,000$. The red arm was equipped with the R1200R 
grating and the {\it Red+} CCD, which gives similar coverage and dispersion. Two exposures, centred on 5600 and 6800\:\AA{} were taken. 
These spectra are listed at the bottom of Tab.~\ref{log}, as $\sharp$12 and $\sharp$13, but they were not 
used for radial velocity determination because of the lower resolution and lack of radial velocity templates.

Photometric data were obtained with two 8-inch aperture telescopes, a Meade LX200 and a Vixen VISAC, with focal ratios of f/6.3 and f/9,
respectively. The observing locations have coordinates: $36\degr37\arcmin31\arcsec$ North, $6\degr21\arcmin43\arcsec$ West and $38\degr26\arcmin23\farcs7$ North, $0\degr26\arcmin15\farcs7$ West, respectively. Observations were made through Johnson $B$ and $V$ filters, with uniform 120-s exposures (though the two filters were used 
on different nights). We reduced the data with the standard commercial 
software packages AIP4Win and Mira Pro.  We followed the standard procedures for bias and flat corrections. A total of 1138 
photometric $V$-band points were registered, randomly distributed in night series between JD 2454875 (12/2/2009) and 
2457048 (26/1/2014). Later 780 $B$-band photometric points were observed between 2015, February 1  and March 5. Photometric values 
were derived differentially with respect to two reference stars with similar colours in the same frame: UCAC4~452-022165  
($V= 11.75$~mag in APASS) and  UCAC4~452-022241 ($V=11.45$~mag in APASS). The former is Star 14 in Dolidze~25, using the 
WEBDA numbering system, and is likely a foreground late-B star. The latter is Star 17 in Dolidze~25, a cluster member 
with spectral type O7\,Vz (Paper~I).

\begin{figure}
   \centering
   \resizebox{\columnwidth}{!}{\includegraphics[clip]{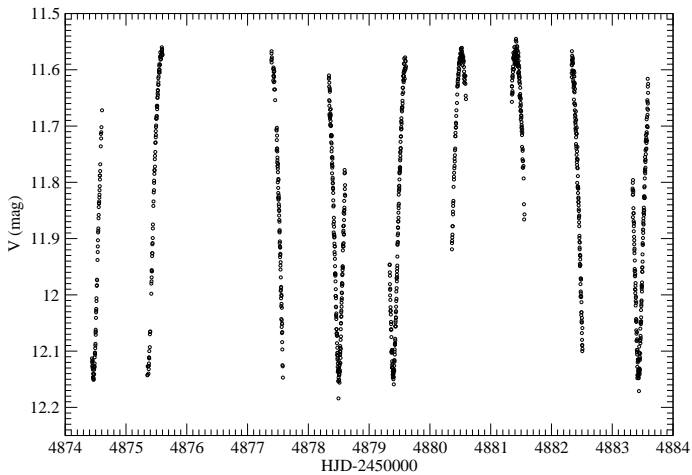}}
      \caption{Photometric light curve for the $V$ filter, showing the amplitude of the modulation.\label{phot_gu_V}}
   \end{figure}

\begin{figure}
   \centering
   \resizebox{\columnwidth}{!}{\includegraphics[clip]{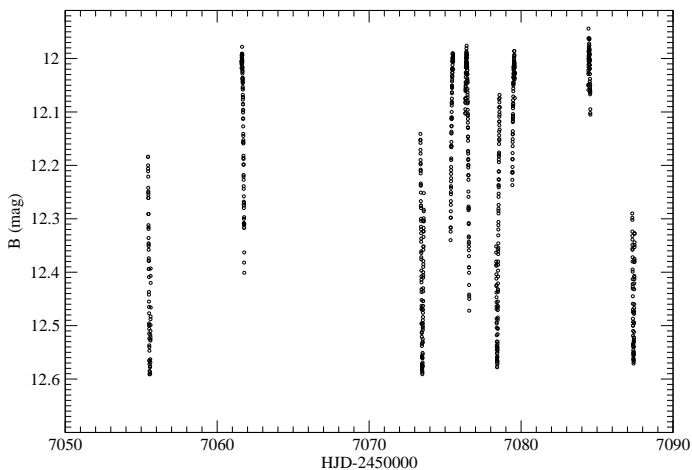}}
      \caption{Photometric light curve for the $B$ filter, showing a very similar amplitude of the modulation.\label{phot_gu_B}}
   \end{figure}

The two lightcurves are displayed 
in Figures.~\ref{phot_gu_V} and~\ref{phot_gu_B}. Strong photometric variability with an amplitude of approximately 0.64~mag 
is apparent for both filters.
 
Even though photometry for the system is available in the CoRoT database, these data are derived from a white light curve, which is the sum of the three chromatic light curves from the CoRoT three-color photometry \citep{auve2009}. The three CoRoT channels (red, green, and blue) do not represent any standard photometric system, and therefore their response to extinction is unknown. Thus, it was felt necessary to obtain new lightcurves in standard Johnson filters, so that a distance to the system could be calculated. Given the quality of our own photometric series, we do not include the CoRoT data in our analysis.

\section{Spectroscopic analysis: radial velocity determination}

\begin{figure}
   \centering
\resizebox{\columnwidth}{!}{\includegraphics[clip]{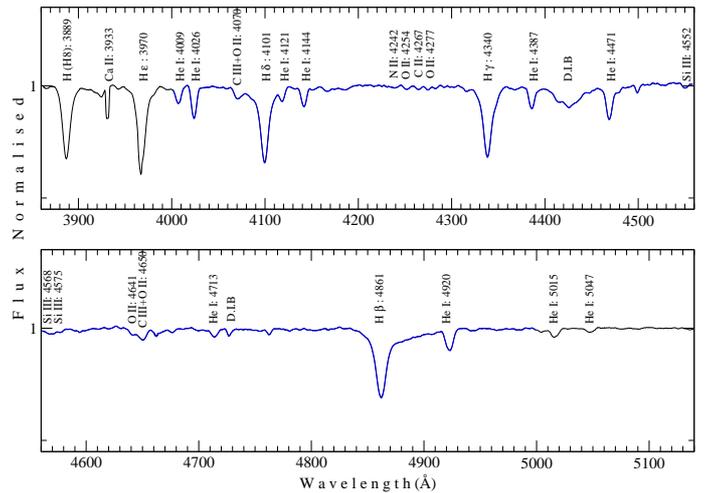}}
      \vspace*{-1.8cm}
      \caption{Classification spectrum of GU~Mon. Labels identify the most characteristic spectral lines. The region marked in blue 
      (between 4000 and 5000\,\AA) represents the spectral range selected for the cross correlation. D.I.B indicates the main
      diffuse interstellar bands.\label{spec}}
   \end{figure}

For spectral classification of GU~Mon, we used the ISIS spectra, which have a much higher signal-to-noise ratio. 
The spectrum is displayed in Fig.~\ref{spec}. We based our classification on the criteria of horizontal classification by \cite{walb1990}. Given the complete absence of \ion{He}{ii} lines, we used as comparison the spectrum of the B1\,V star  $\omega^{1}$~Sco of 
the OB spectral atlas. The spectra are very similar. We can notice the weakness of Si\,{\sc iv}~4089\AA\ and the stronger 
Si\,{\sc iii} triplet (4552\,--\,68\,--\,75\,\AA). We can also notice O\,{\sc ii} lines (perhaps blended with C\,{\sc iii}) at 4070\,--\,72\,\AA\ 
and 4650\AA. The spectrum is dominated by He\,{\sc i} lines (e.g. 4009\,\AA, 4026\,\AA, 4121\,\AA, 4144\,\AA,
 4387\,\AA, 4471\,\AA\ and 4713\,\AA). Even at this resolution, GU~Mon is recognisably a double-lined spectroscopic binary. In the higher resolution FIES spectra, double lines are seen in all the observations, except spectrum $\sharp$8, which corresponds to a phase near eclipse. 
The interstellar H and K lines of \ion{Ca}{ii} can be seen as very narrow features at 3934 and 3969\,\AA\,(see Fig.~\ref{spec}).
  
The cross-correlation method is a widespread and reliable technique to derive radial velocities of the two components of composite 
spectra. In the case of OB stars where spectral lines are very wide due to high rotation, it is the best method to determine accurate 
radial velocities. We used the TODCOR technique based on cross correlation in two dimensions \citep{zuck1994}, which should reduce the
problems caused by blending effects \citep{sout2007}. The observational 
spectrum is cross correlated with a combination of two 
templates simultaneously. TODCOR is specially useful for the analysis of binary spectra with small velocity differences. Before 
computing TODCOR, we had to choose suitable templates. The templates selected were two synthetic spectra derived from the atmosphere 
model fit to the observed spectra, obtained using the {\sc fastwind} code \citep{puls2005,sant1997}. The spectral range between
4000--5000\,\AA\ was chosen for the analysis, because it includes the main He\,{\sc i} lines (4009, 4026, 4121, 4144, 4387, 4471, 4713 and 4920\,\AA)
as well as H$\delta$, H$\gamma$ and H$\beta$. Although \cite{ande1980} conclude that the use of the Balmer lines in the cross-correlation could lead to an under-estimation of the mass of up to 40\% compared to a solution with only the He\,{\sc i} lines, we have used all the atmospheric lines in the spectral range chosen. A solution leaving out the Balmer lines results in a slightly higher mass for the stars, by about 4\%, but much higher standard deviations in the fit to the radial velocity (typically $\sigma\approx24\:$km\,s$^{-1}$.)

Spectrum modelling was performed using a similar strategy as described in \cite{lore2014} and 
\cite{simo2015}. In brief, the stellar parameters were obtained from a by-eye 
comparison of the H and \ion{He}{i}/\ion{He}{ii} profiles from the grid of  {\sc fastwind} synthetic spectra described in 
\citet[these are spectra with $Z=Z_{\sun}$]{simo2011}, and the spectrum $\sharp$6 of GU\,Mon (see numbering in Tab.~\ref{log}). This spectrum was chosen because the two components are clearly separated, and the wings of the Balmer lines can be used to determine the effective gravity with accuracy. Its S/N ratio is higher than those of other spectra at similar orbital phases.
In the analysis process, the synthetic spectra were shifted to the radial velocities indicated in Tab.~\ref{rv_res} and 
convolved with a projected rotational velocity\footnote{Given the quality of the spectra, we 
were only able to roughly constrain the projected rotational velocity between 250 and $300\:{\rm km}\,{\rm s}^{-1}$ for each component. The rotational velocities  derived from model fitting are susceptible to two possible causes of inaccuracy.  On the one hand, the low signal-to-noise of some of the spectral lines can result in inaccurate profile fits. Secondly, the Struve-Sahade effect \citep{stru1937,saha1959}, i.e. the strengthening of the profile corresponding to the component of a spectroscopic binary that is approaching the observer, can contribute to larger uncertainties. This strengthening could perhaps induce differences in the derivation of the projected rotational velocities of each star. In the end, we decided to fix $v\,\sin\,i$ in both stars to $270\:{\rm km}\,{\rm s}^{-1}$, because we checked by eye that this value seemed to reproduce better most of the spectra (and not only the spectrum used for the fit). The convenience of this choice is supported by the results of the combined analysis of the radial velocity and photometric light curve, which gives a perfectly compatible value (see Tab.~\ref{par}).} ($v\,\sin\,i$) of $270\:{\rm km}\,{\rm s}^{-1}$. We found that both stars are equally contributing 
to the spectrum and have very similar effective temperatures and gravities 
($T_{\rm eff}$=28\,000 $\pm$ 2\,000~K, $\log\,g$ = 4.1 $\pm$ 0.2 dex). In addition, a normal Helium abundance ($Y_{\textrm{He}}$=0.10) and a low 
value of the wind strength parameter 
\footnote{$Q=\frac{\dot{M}}{\left( R\,v_{\infty} \right)^{1.5}}$} \citep{puls1996} 
were needed to result in a good fit in the two components.

\begin{figure}
   \centering
\resizebox{\columnwidth}{!}{\includegraphics[clip]{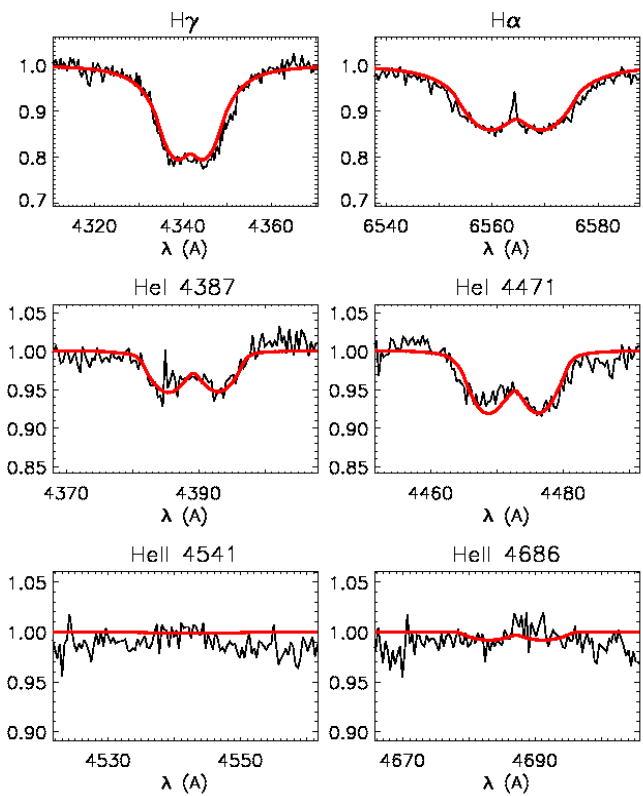}}
      \caption {Some representative H, He\,{\sc i} and He\,{\sc ii} lines in the spectrum of 
      GU~Mon ($\sharp$6 in Tab.~\ref{log}) in used to derive the rotational velocities 
      and stellar parameters of both components. The red solid line is the combined best fitting {\sc fastwind}
      synthetic spectrum.\label{fastwind}}
   \end{figure}

In Fig.~\ref{fastwind}, we show the quality of the fitting to the most relevant H, He\,{\sc i} and He\,{\sc ii} diagnostic line-profiles. We note that,
while the Si\,{\sc iv}/Si\,{\sc iii} ionization balance is better suited to constrain the effective temperature in the case of 
B0\,--\,B2 stars \citep[e.g.][]{simo2010}, the high $v\,\sin\,i$ of both components of GU~Mon, combined with the
low signal-to-noise of the spectrum, did not allow us to use these weaker lines; hence, the relatively large uncertainty 
in the derived temperatures.

The radial velocities of the spectra observed were obtained by applying the TODCOR technique with the two templates mentioned. 
Spectra and templates were rebinned to a length of 10\,000 bins. The centre of the search area was fixed at zero $\:{\rm km}\,
{\rm s}^{-1}$, because 
all the spectra were corrected to their heliocentric velocities. The apodization factor was set to 5\%~of the edges of 
the spectra. Radial velocities and residuals  
are presented in Tab.~\ref{rv_res}, which will be commented in the next section.

\section{Combined analysis: photometry and spectroscopy}

To derive the orbital parameters of GU~Mon, we have performed light curve (LC) and radial velocity curve (RVC) analysis 
together. We have used a program based on the generalised Wilson-Devinney (WD) code \citep{wils1971} in its 2010 version, assuming 
the Roche model, the most accurate approximations based on the principle of equipotential surfaces. LCs and RVC models are 
obtained by fixing some parameters and computing the rest by using differential corrections until free parameter adjustment of LCs and RVC is reached according to the least-squares criterion. 
We have considered that GU~Mon is a contact system, in which case circularisation and synchronisation are acceptable. We have chosen 
mode 1 of the WD code, corresponding to overcontact binaries. In this case, 
the surface potentials are the same for both stars ($\Omega_{1}=\Omega_{2}$). The radiative model for both components of the 
binary system is an atmosphere model by \cite{kuru1993}. The surface is divided into a grid of $40 \times 40$ elements for every star. 
To improve the convergence of the solution, 
we chose symmetrical derivatives \citep{wils1976}. The code can apply the detailed reflection model of \cite{wils1990}, 
a treatment especially recommended with overcontact binaries. We have also considered proximity effects on both stars. 
A square root limb-darkening law was applied during the process, as it is an order of magnitude more precise than the linear 
law \citep{hamm1993}. The bolometric albedos of both components were fixed $A_1=A_2=1$, because the atmospheres are expected to be 
in radiative equilibrium \citep{vonz1924}. Because of
local energy conservation, this also implies gravity brightening exponents $g_1=g_2=1$. Other constraints applied are described in 
mode 1 of the WD code.

A preliminary analysis of the spectroscopy provided the initial values for parameters such as period, zero point of ephemeris, and so 
on. The stellar parameters, such as temperatures, are derived from the atmosphere model.
The temperature ratio is not an adjustable parameter during the convergence process. This is a compulsory constraint due to the morphology 
of the binary star, as both components are sharing a volume of their Roche lobes, and thermal contact is accepted.

The process to convergence of all free parameters is iterative and simultaneous for all observables analysed, i.e. the radial 
velocity curves and the light curves. The criterion for convergence adopted is that, for three consecutive iterations, all 
adjustable parameters must be within two standard deviations. Once convergence is reached, five solutions are derived 
by varying the parameters within the standard deviation and fitting the observations again. We choose the fit with the smallest 
dispersion as a final solution. 

\subsection{Light Curves}

\begin{figure}
   \centering
   \resizebox{\columnwidth}{!}{\includegraphics[clip]{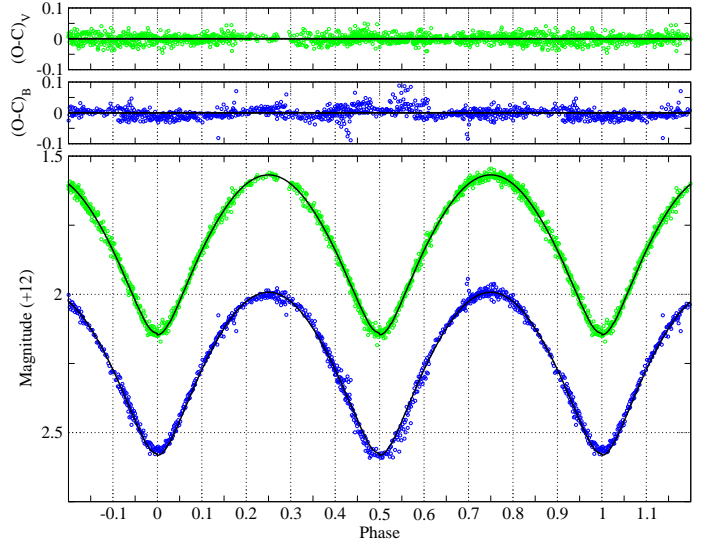}}
      \caption{Light curve model fitted to the observational data in $V$ filter (green) and $B$ filter (blue). 
      Residuals are displayed in the top panel. \label{lc_res}}
   \end{figure}

We classify the lightcurve of GU~Mon  phenomenologically. GU~Mon as an eclipsing variable of the EW type. The differences between star~1 and star~2 
are so small that they are almost indistinguishable. To quantify this similarity, we explored if there was any difference between the two 
photometric minima. We averaged all photometric data points between $\phi=0.99$ and~$0.01$, and all points between $\phi=0.49$ 
and~$0.51$, obtaining a difference of 6 tenths of a millimagnitude for the $V$ filter and 2 millimagnitudes for the $B$ filter. 
In both cases, this is smaller than the intrinsic dispersion ($\sigma$) of the corresponding light curve. Accordingly, it is not 
possible to decide which is the primary minimum and which the secondary minimum. In this case, any of the
eclipses can be considered as primary, and so the zero time of ephemeris can be chosen with half a cycle difference, as has happened 
in the literature \citep{zejd2002}. There is no plateau between the eclipses. The light curve exhibits a continuous and monotone 
shape along the cycle. This shape of the model lightcurve confirms that both stars are filling and sharing their Roche lobes. We 
have made sure that the zero phase of the LCs fit corresponds to the zero phase of the orbital solution. As a further check, we also determined the period of 
the system from the LCs and the RVC independently, obtaining the same value for both methods. The difference between the 
photometric and the spectroscopic period is less than one second, and thus they are the same within errors. As mentioned above, the temperature was assumed to be the same for both stars and fixed to 28\,000~K. The luminosity was considered a free parameter, together with the surface potential of the stars (see Tab.~\ref{par}). LCs for both filters are displayed in Fig.~\ref{lc_res} with the model of light curve and residuals yielding a $\sigma_{V}\approx 0.013$ and  $\sigma_{B}\approx 0.019$. Almost all the residuals are under 0.05~mag for the $V$ filter and under 0.09~mag for the $B$ filter, showing in any case an excellent fit.

The linear ephemeris equation, where the epoch of successive times of primary-eclipse minima (phase zero), $T_{{\rm min}}$, is calculated 
from the period and zero time ephemeris as: 

\begin{multline*}
 T_{\rm min} = {\rm HJD}~(2454874{.}46611\pm0{.}00014)\\
 + (0^{\rm d}{.}896640\pm0^{\rm d}{.}000007)\times E \:\,
\end{multline*}

where $E$ is the integer value of the number of orbital cycles.

\subsection{Radial velocity curve}

\begin{figure}
   \centering
    \resizebox{\columnwidth}{!}{\includegraphics[clip]{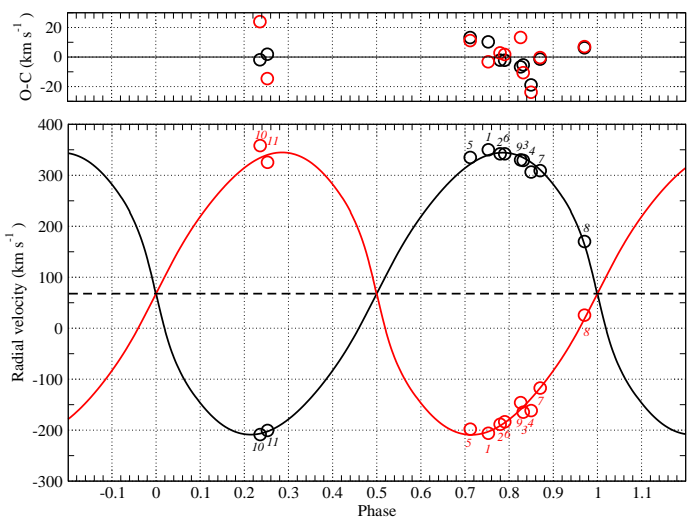}}
      \caption{Radial velocity curves fitted to the observational data and displayed against orbital phase 
(black line: star 1; red line: star 2).
The dashed line corresponds to the systemic velocity.
The residuals are displayed in the top panel. \label{rvc-res}}
   \end{figure}

Simultaneously to the LCs analysis, we derived some orbital parameters such as systemic velocity,  mass ratios and semimajor axis 
(see Tab.~\ref{par}) by fitting the RVC, which is displayed in Fig.~\ref{rvc-res} (bottom panel). Both stars have the same semi-amplitude 
of the velocity, and again we cannot assign the classical designation
of primary and secondary star: both have the same mass. Therefore we stick to the designations star 1 and star 2, as with the LCs analysis. 
In Fig.~\ref{rvc-res}, each radial velocity is marked with the number corresponding to the spectrum in Tab.~\ref{log}. Even though the 
eccentricity is zero, we can distinguish a significant deviation from a sinusoidal shape. This is likely due to the distortion 
created in the stars by the proximity effects. Residuals are displayed in Fig.~\ref{rvc-res} (top). The fit to the RVC of star~1 
gives smaller residuals ($\sigma\approx8\:$km\,s$^{-1}$) than the fit obtained for star~2 ($\sigma\approx$ 13\:km\,s$^{-1}$). As expected, the fit shows that both stars share the same systemic velocity. 

We have sorted the radial velocities according to the orbital phase (see Tab.~\ref{rv_res} and Fig.~\ref{rvc-res}). Residuals are in all cases below 
25\,km\:s$^{-1}$. Taking into account the proximity of the stars, and the width of the spectral lines due to high rotational velocity, the model fit can be considered very good.

\begin{table}
\caption{Radial velocities and residuals for 
both components of GU~Mon, sorted by orbital phase. \label{rv_res}}      
\centering 
\scalebox{0.90}{   
\begin{tabular}{c c c c c c}        
\hline\hline   
\noalign{\smallskip}        
Number&Phase & RV$_{\rm 1}$ &O-C$_{\rm 1}$& RV$_{2}$&O-C$_{\rm 2}$\\   
   &  &(km\:s$^{-1}$)&(km\:s$^{-1}$) &(km\:s$^{-1}$)&(km\:s$^{-1}$)\\
\noalign{\smallskip}      
\hline
\noalign{\smallskip}      
10 &0.236 &  $-209.0$	&  $-1.9$ &  358.0    & 24.0\\ 
11 &0.253 &  $-201.0$	&   1.9   &  326.0    &$-15.0$\\ 
 5 &0.712 &	335.0	&  13.0   & $-198.0$  & 11.0\\ 
 1 &0.753 &	350.0	&  10.0   & $-206.0$  & $-3.0$\\ 
 2 &0.780 &	342.0	&  $-2.1$ & $-188.0$  &  2.8\\ 
 6 &0.790 &	342.0	&  $-2.2$ & $-184.0$  &  1.8\\ 
 9 &0.826 &	330.0	&  $-7.0$ & $-146.0$  & 13.0\\ 
 3 &0.832 &	329.0	&  $-5.0$ & $-165.0$  &$-11.0$\\ 
 4 &0.850 &	306.0	& $-19.0$ & $-162.0$  &$-24.0$\\ 
 7 &0.870 &	309.0	&  $-1.4$ & $-117.0$  & $-0.4$\\ 
 8 &0.971 &	170.0	&   6.0   &   26.0    &  7.0\\ 
\noalign{\smallskip}      
 \hline                        
\end{tabular}}
\end{table}

\subsection{Results}

We have derived stellar parameters for both components of GU~Mon (listed in Tab.~\ref{par}), using
simultaneous solutions for the LCs and RVC, as explained in the previous section. The uncertainties in the parameters were computed 
by propagating the uncertainties in the observed quantities. The eccentricity was assumed to be zero, as the review of \cite{torr2010} 
shows that no measurable eccentricity has been found for any system with a period below 1.5~d. Under these conditions, we can expect the 
system to have circularised. Based on the time scales for synchronisation and circularisation given by \cite{zahn1977}, we can expect 
$t_{\textrm{sync}}\sim {10}^{5}$~yr and $t_{\textrm{circ}}\sim {10}^{7}$~yr for the stellar masses found in GU~Mon. These time scales 
are very strongly dependent on the ratio ($a$/$R$), the separation divided by the stellar equatorial radius. For very short periods 
and $R$/$a\sim0.5$, as we see in GU~Mon, we can reasonably expect a synchronous rotation and a circular orbit, as we have verified 
in the results obtained. 
The projected rotational velocity obtained from the modelling is $268\pm5\:{\rm km}\,{\rm s}^{-1}$, while the semi-amplitude of the 
orbital velocity is $276\pm6\:{\rm km}\,{\rm s}^{-1}$. Given that the spectral modelling supports a rotational velocity for both components in the 250\,--$300\:{\rm km}\,{\rm s}^{-1}$ (with a favoured value around $270\:{\rm km}\,{\rm s}^{-1}$), the assumption of synchronisation seems to be fully supported, and we conclude that both stars are tidally locked. In fact, if we study the variation of the equatorial radii due to proximity effects, we can
deduce projected rotational velocities in a range from 245\:km\:s$^{-1}$ (corresponding at equatorial angle
of $76\fdg2$ and its explementary angle of $283\fdg8$, both measured from the semimajor axis) 
until 355\:km\:s$^{-1}$, close to the point radius, unapproachable for the calculation because both
components share their gaseous envelopes.

\begin{figure}[!ht]
\begin{center}
\resizebox{\columnwidth}{!}{\includegraphics[clip]{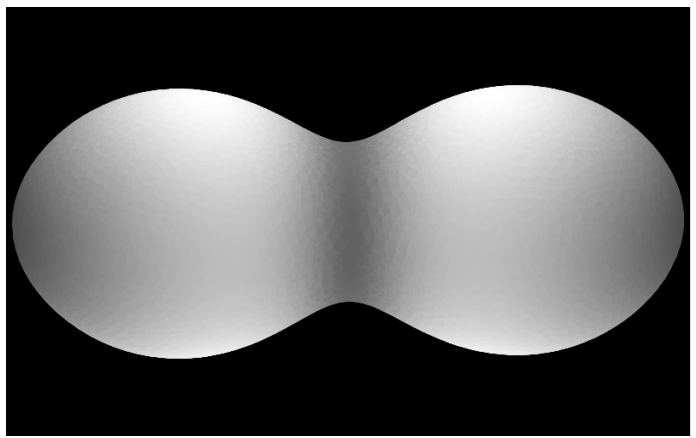}}
\end{center}
\caption{Representative drawing of GU~Mon to scale at quadrature phase, created with the {\it PHOEBE} 
2.0-alpha code via the Python interface. \label{quad_gumon}}
\end{figure}

GU~Mon is an overcontact
system, because the effective radius of the Roche lobe \citep{eggl1983} is $3.63\,R_{\sun}$, and so the radii of both stars are 
larger than the effective radius of the Roche lobe (see Fig.~\ref{quad_gumon}). The fill-out factor 
(here calculated according to the description on page~110 of \citealt{kall2009}) expresses the degree of contact for the binary system. 
If we consider the equatorial radius pointing to the other component, the fill-out factor of GU~Mon is $\sim0.8$. A fill-out factor 
equal to unity would imply that the normalised surface potential is equal to the outer normalised surface
potential passing through the Lagrangian point $L_2$. Therefore the two stars are not only touching, but sharing a substantial fraction of their envelopes. In view of this, we can speculate about the possibility that the indistinguishability of the two stars is a consequence of mass transfer, which has led to an object that is, in many senses, a single star with two cores. 

\begin{table}
\caption{Stellar parameters derived from the combined analysis of the 
radial velocity and photometric light curve. \label{par}}    
\centering 
\scalebox{0.9}{   
\begin{tabular}{l c c }        
\hline\hline
\noalign{\smallskip}           
&Star~1& Star~2\\
\noalign{\smallskip}  
\hline
\noalign{\smallskip}  
Orbital period (day)&\multicolumn{2}{c}{0.896640 $\pm$ 0.000007}\\
Zero point of ephemeris (HJD)&\multicolumn{2}{c}{2454874.46611 $\pm$ 0.00014}\\
Eccentricity&\multicolumn{2}{c}{0 (assumed)}\\
Inclination ($^{\circ}$)&\multicolumn{2}{c}{72.34 $\pm$ 0.09}\\
Longitude of periastron ($^{\circ}$)&90&270\\
Systemic velocity (km s$^{-1}$)&\multicolumn{2}{c}{68 $\pm$ 5}\\
Semi-amplitude of velocity (km s$^{-1}$)&276$\pm$6&276 $\pm$ 6\\
Semimajor axis (R$_{\sun}$)& \multicolumn{2}{c}{10.25 $\pm$ 0.22}\\
Surface normalised potential&\multicolumn{2}{c}{3.340 $\pm$ 0.006}\\
Mass ($M_{\sun}$)&9.0 $\pm$ 0.6&9.0 $\pm$ 0.6\\
Mass ratio ($M_{2}/M_{1}$)&\multicolumn{2}{c}{0.999 $\pm$ 0.004}\\
Mean equatorial radius ($R_{\sun}$)&4.98 $\pm$ 0.12&4.98 $\pm$ 0.12\\
Polar radius ($R_{\sun}$)&4.24 $\pm$ 0.10&4.24 $\pm$ 0.10\\
Side radius ($R_{\sun}$)&4.60 $\pm$ 0.12&4.60 $\pm$ 0.12\\
Back radius ($R_{\sun}$)&5.48 $\pm$ 0.16&5.48 $\pm$ 0.16\\
Projected rotational velocity\footnotemark[1] (km s$^{-1}$)&268 $\pm$ 5&268 $\pm$ 5\\
Surface effective gravity\footnotemark[2] ($\log\,g$) &4.07 $\pm$ 0.04&4.07 $\pm$ 0.04\\
Luminosity ratio ($L_{2}/L_{1}$)&\multicolumn{2}{c}{0.9999 $\pm$ 0.0015}\\
\hline 
\noalign{\smallskip}  
\end{tabular}}
\scriptsize{
\tablefoottext{1}{calculated from the mean equatorial radius}
\tablefoottext{2}{calculated from the side radius}}
\end{table}

Eclipsing binaries allow the derivation of geometrical distances to the systems, which can be very precise \citep[e.g.][]{sout2004, vila2010}. In overcontact binaries, there are many complications, due to the interaction and 
geometrical distortion, but a direct distance estimate is still possible from the stellar parameters. We used the effective
temperature 28\,000~K, obtained from the atmosphere model fit to compute a distance-dependent flux, in standard physical units, 
including those proximity effects supported by the model. This procedure, though approximate, is more accurate than the simple 
use of the mean radii. We follow the procedure described in \cite{vila2010}.
The parameters used  are showed in Tab.~\ref{distance}, together with our distance calculation, $3.8\pm0.5$~kpc. The large error is mainly due to the relatively large uncertainties in the extinction and the bolometric correction. Indeed, the uncertainty in the extinction is underestimated, because we have simply assumed a standard extinction law, as this seemed enough to describe the reddening of most cluster members \citepalias{negu2015}.

\begin{table}
\caption{Parameters resulting from the direct estimation distance of GU~Mon. \label{distance}}    
\centering 
\scalebox{0.9}{   
\begin{tabular}{l c c }        
\hline\hline
\noalign{\smallskip} 
Binary properties&&\\
\noalign{\smallskip} 
\hline
\noalign{\smallskip} 
$E(B-V)$ (mag)  & $0.71\pm0.03$&\\
$A_V$           & $2.2\pm0.23$&\\
$M_V$ (mag)     &$-3.56\pm0.15$&\\
$(V_0-M_V)$ (mag)&$12.92\pm0.28$&\\
distance (pc)    &$ 3800\pm500$&\\
\noalign{\smallskip} 
\hline
\noalign{\smallskip} 
Component properties& star 1&star 2\\
\noalign{\smallskip} 
\hline
\noalign{\smallskip} 
$ M_V$    (mag) &      $-2.81\pm0.15$  &  $-2.81\pm0.15$\\
$(B-V)_0$ (mag)&     $-0.28\pm0.01$  &    $-0.28\pm0.01$\\
\noalign{\smallskip} 
\hline 
\end{tabular}}
\end{table}

\section{Discussion}

We have calculated full orbital and stellar parameters for GU~Mon. Our model fits to the light curves and the RVC are 
very good given the quality of the spectroscopic data. All the stellar parameters derived from the atmosphere model 
agree with those deduced from the orbital analysis. The projected rotational velocity has been discussed in the previous 
section in the context of circularisation. In the case of surface gravity, the agreement is perfect. The atmosphere model 
gives $\log\,g_{\rm}$ $= 4.1 \pm 0.2$ and the orbital analysis, $\log\,g_{\rm}$ $= 4.07 \pm 0.04$. Therefore we can be sure 
that the stars are on the main sequence, and not far from the ZAMS. Finally, the direct distance estimation to GU~Mon is 
$3.8\pm0.5$~kpc, compatible within errors with the distance derived to the cluster Dolidze~25 of $4.5\pm0.5$~kpc, based 
on several indicators \citepalias{negu2015}. The systemic velocity of GU~Mon is $68\pm5\:\textrm{km}\,\textrm{s}^{-1}$ 
in the heliocentric system, which corresponds to $v_{\textrm{LSR}}=54\pm5\:\textrm{km}\,\textrm{s}^{-1}$. 
As shown in \citetalias{negu2015}, other members of the cluster show radial velocities between 
$45\pm3\:\textrm{km}\,\textrm{s}^{-1}$ and $48\pm3\:\textrm{km}\,\textrm{s}^{-1}$. The radial velocity 
of GU~Mon is thus fully compatible with cluster membership.

\begin{table}[!ht]
\centering
\begin{minipage}{\columnwidth}
\caption{Close binaries with early-type stars on the main sequence
  and orbital period shorter than one day (extracted from \citealt{polu2004}).\label{bin_1dia}}
\resizebox{9cm}{!} {  
\begin{tabular}{lccc}
\hline
\hline
\noalign{\smallskip}
Name &Period (d)&Spectral type&Mass (M$_{\odot}$)\\
\noalign{\smallskip}
\hline
\noalign{\smallskip}
BH Cen&0.792&B3+B3&9.4 $\pm$ 5.4 + 7.9 $\pm$ 5.4 (1)\\
V593 Cen&0.755&B1\,Vn&\textrm{unknown}\\
EM Cep&0.806&B0.5\,V+B1\,Ve&10.51 + 9.46 (2)\\
BR Mus&0.798&B3&\textrm{unknown}\\
V701 Sco&0.762&B1\,V-B1.5\,V&10.3 + 10.2 (3)\\
CT Tau&0.667&B2+B2&\textrm{unknown}\\
\noalign{\smallskip}
\hline
\end{tabular}
}
\begin{list}{}{}
\item[] \scriptsize{ References: (1) \citet{leun1984}, (2) \citet{hohl2010}; (3) \citet{polu2004}}.
\end{list}
\end{minipage}
\end{table}

The masses of the two components of GU~Mon are broadly compatible with their spectral type, but slightly lower than expected.
The solution obtained by applying TODCOR while masking the Balmer lines gives masses of $9.4\:M_{\sun}$. This value lies within the uncertainty calculated (see Tab.~\ref{par}). Given the much higher residuals in the radial velocity curve fit, we decided to stick to the result obtained with the whole set of lines.
For comparison, we can consider some other binaries with B1\,V components. The overcontact system V701~Sco, which will be 
discussed below, contains two identical stars with mass $\approx10.2\:M_{\sun}$ (Tab.~\ref{bin_1dia}). Among detached binaries, 
one of the best examples is DW~Car, a member of the very young open cluster Collinder~228, whose components have been both 
classified as B1\,V. With an orbital period of $1.3\:$d, they have masses of $11.34\pm0.12$ and $10.63\pm0.14\:M_{\sun}$ 
\citep{sout2007,clau2007}. This object is almost certainly a pre-contact binary, and both components seem to be still very 
compact, as in  GU~Mon. The cooler component of V578~Mon, in the very young cluster NGC~2244, has a temperature and radius similar 
to those of GU~Mon, and a mass of $10.3\:M_{\sun}$ \citep{hens2000}. This is also a pre-contact system. Finally, V~Pup is a 
post-mass transfer system, where the B1\,V primary (very likely the mass gainer) has a mass of $12.5\:M_{\sun}$ \citep{stick1998}. 

\begin{figure}
   \centering
\resizebox{\columnwidth}{!}{\includegraphics[clip, angle=90]{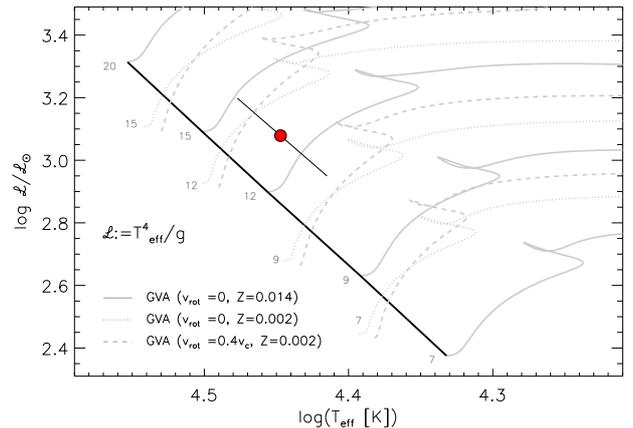}}
      \caption {Location of the two "twin" components of GU~Mon in the spectroscopic HR diagram, including the uncertainties in $T_{\textrm{eff}}$ and $\log {\mathscr L}$. Three sets of tracks from the Geneva group
 are overploted to illustrate the effects of metallicity and rotation. Only the ZAMS for the non-rotating models at solar metallicity is shown.\label{shrd}}
\end{figure}

The relatively low masses for the components of GU~Mon may be partially explained by the low metallicity of Do~25. With an oxygen abundance $-0.4$~dex below solar, Do~25 has a metallicity lower than the LMC. Stars at lower metallicity are known 
to be more compact and hotter at a given mass. To check this effect, in Fig.~\ref{shrd} we plot the position of the components of GU~Mon in the spectroscopic HR diagram \citep{lang2014}, together with theoretical isochrones from the Geneva group corresponding to solar and SMC-like metallicities \citep{ekst2012,geor2013}. The metallicity of Do~25 falls approximately between the two, somewhat closer to the SMC. The solar metallicity tracks would indicate masses of $\approx13\pm2\:M_{\sun}$ for each of the stars, higher than any of the B1\,V stars in eclipsing binaries, even though the values of $T_{\textrm{eff}}$ and $\log\,g$ that we find for GU~Mon can be considered typical of the spectral type. This is probably a reflection of the well-known discrepancy between evolutionary masses and other determinations.
For SMC metallicity, the mass of the stars would be $\approx11\pm2\:M_{\sun}$, still higher than the dynamical values, but now compatible within errors. 

On the other hand, the fact that numerous metallic lines are clearly seen in the spectrum even at this low metallicity means that the spectral type cannot be later than B1. Several weak \ion{O}{ii} lines that have disappeared by B2\,V are clearly seen in Fig.~\ref{spec} in spite of the high rotational velocity. Even at B1.5\,V, the \ion{C}{ii}~4267\:\AA{} line should be rather stronger than the lines surrounding it, unless Carbon is severely depleted. But the spectrum shows no evidence for the Nitrogen enhancement that generally accompanies Carbon depletion, and that could be a signature of chemical mixing. Therefore we have to conclude that the components of GU~Mon are two ``normal'' low-metallicity B1\,V stars. A possible reason for their anomalous position in the sHR diagram is the inability of theoretical models of single stars to reproduce the properties of stars in close contact. As of now, there are no theoretical models for high-mass binaries able to predict the physical properties of the stars during the overcontact phase. In support of this possibility, we have to note that all the B1\,V stars with dynamical masses listed in the preceding paragraphs have lower masses than predicted by the solar-metallicity tracks. According to \citet{alme2015}, this also implies that a realistic treatment of heat transfer during the contact phase is missing. If so, the masses derived from our analysis could be less accurate than what the formal errors of the WD code would indicate or perhaps the stars could be hotter than corresponds to their mass because of the common envelope.

Our results show that GU~Mon is in an advanced stage of interaction. Even so, its parameters are not very unusual. 
In Table~\ref{bin_1dia}, we list known binaries whose components are OB stars on the main sequence with a period shorter 
than a day, as in GU~Mon. For three of them, accurate parameters have not been derived. The components of BH~Cen have 
spectral types later than those of GU~Mon, and their masses are not known with accuracy.  BH~Cen belongs 
to the young open cluster IC~2944, where star formation is in process \citep{qian2006}, as in Dolidze~25. Two systems, 
EM~Cep and V701~Sco, are very similar to GU~Mon, having comparable masses. However, we must caution 
that \cite{hild1982} argued
that EM~Cep is not an eclipsing binary at all, but instead a $\beta$~Cep pulsator, with the photometric variability being due to 
pulsations. This hypothesis has not been refuted to date\footnote{We note that this possibility is ruled out for GU~Mon, which clearly shows double lines moving in phase with the photometric lightcurve and becoming a single line at the times of eclipses. Moreover, there are no noticeable variations in the depth of the spectral lines.}. V701~Sco, on the other hand, seems an excellent analogue to GU~Mon. 
It is a member of the very young open cluster NGC~6383 \citep{saha1963, lloy1978}. Again, star formation is still active in this 
cluster, centred on the 3.4~d binary HD~159\,176  \citep[O7\,V((f))+O7\,V((f));][]{lind2007}. Like Do~25, NGC~6383 has an estimated 
age slightly below 3~Ma, and contains many pre-main-sequence stars \citep{rauw2010}. \citet{bell1987} studied the evolutionary state 
of V701~Sco, coming to the conclusion that the system had formed already close to contact, with a mass ratio close to unity. 
\citet{qian2006} go as far as suggesting that V701~Sco formed already in an overcontact configuration. By analogy, GU~Mon, 
that has a similar age and orbital period, should also have formed in a configuration very similar to the present one.

The presence of all these systems in young open clusters suggests that similar systems, with twin components in a contact configuration are quite common. Recent spectroscopic surveys \citep[e.g.][]{sota14} are showing that most of the stars classified as ``n'' or ``nn'' (i.e. with broad lines) in photographic spectra are in reality close binaries with high rotational velocities. However, moderately-high spectral resolution is needed to see the two components, and so many more similar binaries may be hidden in young open clusters.

The evolution of such systems is not clear. Given the very short orbital periods and mass ratios close to unity, slow case A mass transfer is likely to dominate their evolution \citep{pols1994,demink2007}. According to \citet{well2001}, if the two systems fill their Roche lobes while the primary is still burning hydrogen in its core, a merger is the most likely outcome, as the short orbital period is believed to preclude common envelope ejection. Given the age of Do~25 ($\la3\:$Ma), the components of GU~Mon are still quite close to the ZAMS, and they have already filled their respective lobes. As a matter of fact, unless the original orbital period was considerably longer, evolutionary models indicate that they were already in contact at age zero \citep{well2001}. The exact initial period for which merger is unavoidable is not known, because of the extreme difficulty in reproducing the physics of the common envelope phase. Some semi-detached systems with short periods are believed to be post-contact binaries, whose orbital periods are increasing after a minimum was reached during fast case A mass transfer. Examples of this are TU~Mus \citep[$P_{\rm orb}=1.4\:$d, O7\,V+O8\,V;][]{penny2008}, or IU~Aur\footnote{However, we have to note that in both TU~Mus and IU~Aur, there is a fair chance that the binary evolution is driven by a third body in the system \citep{qian2006}.} \citep[$P_{\rm orb}=1.8\:$d, O9.5\,V+B0.5\,V;][]{ozdem2003}. In these cases, the contact phase should be short lived \citep[and references therein]{qian2013}. In contrast, the detection of several systems with early B components in a contact configuration with orbital periods $<1\:$d suggests that they are in a long-lived phase that will lead to merger. The only way in which a merger could be avoided is if rotational mixing leads to homogeneous evolution, forcing the two stars to become hotter and more compact. However, the observed rotational velocity for the components of GU~Mon, about $270\:{\rm km}\,{\rm s}^{-1}$, seems far too small to lead to homogeneous evolution. In this aspect, it is similar to the much more massive system MY~Cam \citep{lore2014}. In addition, the lack of nitrogen enhancement does not support rotationally-induced chemical mixing, though this could be hidden if the common envelope is effectively detached from the cores. In all, however, the current evidence strongly supports a future evolution leading to merger and the formation of a blue straggler in Do~25.

\section{Conclusions}

Our analysis has shown that GU~Mon is an overcontact eclipsing binary with almost identical components. The two stars present parameters typical of main sequence stars of spectral type B1\,V. Given the age of the cluster Do~25, of which they are members, they cannot be very far away from the ZAMS. The derivation of accurate stellar parameters is hindered by the shallowness of all lines, caused by the high rotational velocity of the stars. The present-day rotational velocity of the stars, however, is far too low for homogeneous evolution. According to evolutionary models \citep{well2001}, the system should be transferring material slowly and will evolve towards a merger. The result of this merger will probably be a main-sequence star with a mass close to the total mass of the system, which would then be a blue straggler in Do~25. Our modelling shows that, at this stage, the binary already possesses a single atmosphere wrapping both cores. The transfer of material from the inner layers to the common envelope in this type of overcontact binaries has not been studied, and so we cannot obtain information from the stellar interior (for example, we do not know if we should expect evidence of chemical mixing). From the outside, the stellar atmosphere resembles that of a single star but with a very specific morphology, a \textquotedblleft peanut shape \textquotedblright. GU~Mon is not unique, as there seem to be several similar systems, with periods shorter than one day. For all these objects, reversal of the orbital evolution seems unlikely, as no examples of high-mass systems with increasing periods and orbital periods shorter than $\sim1.5$~d are known. Therefore GU~Mon seems to represent a point in evolution between the initial contact and the definitive merger of both components.

\begin{acknowledgements}

Partially based on observations made with the Nordic Optical Telescope, operated by the Nordic Optical Telescope Scientific Association, and the WHT, operated on the  island of La Palma by the Isaac Newton Group, both in the Spanish Observatorio del Roque de Los Muchachos of the Instituto de Astrof\'{\i}sica de Canarias. The WHT observations were taken as part of the service programmes, and we would like to thank the staff astronomers for their diligence.

 This research is partially supported by the Spanish Ministerio de Econom\'{\i}a y Competitividad under 
grants AYA2012-39364-C02-01/02 and AYA2015-68012-C2-1/2, and the European Union. This research has made use of the Simbad service developed at the Centre de Donn\'ees Astronomiques de Strasbourg, France. 
\end{acknowledgements}

\end{document}